\documentclass[apj]{emulateapj}
\shorttitle{Release of the TelFit Code}
\shortauthors{Gullikson et al.}
\usepackage{graphicx}
\usepackage{natbib}
\usepackage[colorlinks,urlcolor=blue,citecolor=blue,linkcolor=blue]{hyperref}
\usepackage{amsmath}
\usepackage[caption=false]{subfig}

\begin{document}
\submitted{Accepted to the Astronomical Journal}
\title{Correcting for Telluric Absorption: Methods, Case Studies, and Release of the TelFit Code}
\author{Kevin Gullikson}
\affiliation{University of Texas, Department of Astronomy}
\author{Sarah Dodson-Robinson}
\affiliation{University of Delaware, Department of Physics and Astronomy}
\author{Adam Kraus}
\affiliation{University of Texas, Department of Astronomy}

\email[Corresponding Author Email: ]{kgulliks@astro.as.utexas.edu}
%

\begin{abstract}
Ground-based astronomical spectra are contaminated by the Earth's atmosphere to varying degrees in all spectral regions. We present a Python code that can accurately fit a model to the telluric absorption spectrum present in astronomical data, with residuals of $\sim 3-5\%$ of the continuum for moderately strong lines. We demonstrate the quality of the correction by fitting the telluric spectrum in a nearly featureless A0V star, HIP 20264, as well as to a series of dwarf M star spectra near the 819 nm sodium doublet. We directly compare the results to an empirical telluric correction of HIP 20264 and find that our model-fitting procedure is at least as good and sometimes more accurate. The telluric correction code, which we make freely available to the astronomical community, can be used as a replacement for telluric standard star observations for many purposes.
\end{abstract}


\maketitle

\section{Introduction}
All ground-based astronomical spectra suffer from contamination by the Earth's atmosphere, which introduces so-called telluric lines into the spectrum. The wavelength of the telluric lines change very slightly with wind along the line of sight and pressure shifts in water lines, and the relative line strength can vary a great deal with the observatory location, weather and the airmass of the observation. The telluric lines must be removed from the spectrum to retrieve many stellar features redward of about 500 nm, which is a nontrivial task.

Often, telluric lines are removed by observing a rapidly rotating hot star (spectral type A or late B) near the same time and airmass as the target star and using it as a template for the telluric spectrum. This approach has several disadvantages: (1) it can be difficult to find a suitable standard star near the same time and airmass as the observation, so approximate corrections are made to the line strengths using Beer's law \citep{Beer1852}; (2) The water vapor content changes much more rapidly than the other absorbing species in the atmosphere, so scaling the whole empirical telluric spectrum is incorrect (3) the standard star has strong hydrogen and helium lines and weak metal lines that can further contaminate the science spectrum; (4) observing a standard star can take a significant amount of precious telescope time, especially for high signal-to-noise work.

A better solution is often to generate a theoretical telluric absorption spectrum from a line list and the observing conditions. Recently, several groups \citep{Seifahrt2011, Bertaux2013, Husser2013, Cotton2013, Gullikson2013} have used the LBLRTM code\footnote{\url{http://www.rtweb.aer.com/lblrtm_description.html}} \citep[][Line By Line Radiative Transfer Model]{Clough2005} for this purpose. However, the interface for the LBLRTM code can be difficult to learn, and can not directly fit an observed spectrum. 

In this paper we describe Telfit\footnote{The TelFit code is available for download from \url{www.as.utexas.edu/~kgulliks/projects.html}}, a code that acts as a wrapper to LBLRTM and allows for easy fitting of the telluric spectrum in astronomical data. We compare the results to those obtained from empiricial telluric correction in optical echelle spectra, and find that model-fitting produces similar telluric line residuals and is in fact better for dim targets or high signal-to-noise ratio work. We only demonstrate the use of TelFit in optical spectra in this work, but note that an earlier version of this code corrected telluric methane absorption near $2.3 \mu$m to $\sim 5\%$ of the continuum level \citep{Gullikson2013}. \cite{Seifahrt2011} provide a case study of telluric modeling in several bands in the near-infrared using a different code, and find similar results.

We describe the observations and data reduction used in this paper in Section \ref{sec:obs}. We then describe the TelFit code and give a brief description of the telluric fitting procedure in Section \ref{sec:tellcorr}. Finally, we demonstrate the use of the TelFit code to correct several optical telluric bands in Section \ref{sec:results}.

\begin{table*}
  \centering
  \begin{tabular}{|ccccccc|}
  \hline
    Star &    R   &  $\rm K_s$ & SpT  & UT date  & Spectrograph & SNR    \\
         &  (mag) & (mag)      &      & yyyymmdd &              & @820 nm \\ 
    \hline \hline
    
    GJ 83.1  &  10.94  &  6.65  &  M4.6  &  20120902  &  MIKE  &  200  \\
    GJ 109 &  9.49  &   5.961   & M3.0 & 20120902 & MIKE        & 219 \\
    GJ 173  &  9.33  &  6.09  &  M1.6  &  20130203  &  MIKE  &  166 \\
    GJ 273  &  8.70  &  4.86  &  M3.6  &  20130204  &  MIKE  &  215\\
    GJ 447  &  9.86  &  5.65  &  M4.0  &  20130203  &  MIKE  &  324 \\
    GJ 581  &  9.46  &  5.84  &  M2.6  &  20130202  &  MIKE  &  336 \\
    GJ 628 &  8.92  &  5.08  &  M3.0  &  20120716  &  MIKE  &  346 \\
    GJ 908  &  8.03  &  5.04  &  M1.0  &  20120717  &  MIKE  &  328 \\
    HD 33793  &  7.90  &  5.05  &  M1.0  & 20130203  &  MIKE  &  315 \\
    HIP 20264  &  5.38  &  5.33 &  A0.0  &  20140301  &  CHIRON  &  167 \\
    HIP 25608  &  5.54  &  5.50  &  A1.0  &  20140301  & CHIRON  &  214 \\
    \hline
  
  \end{tabular}
  \caption{Observations used in this paper. }
  \label{tab:sample}

\end{table*}

\section{Observations and Reduction}
\label{sec:obs}
We observed representative spectra for the early and late-type stars listed in Table \ref{tab:sample}. The A-stars HIP 20264 and HIP 25608 were observed with the CHIRON spectrograph on the 1.5m telescope at Cerro Tololo Interamerican Observatory (CTIO). This spectrograph has a spectral resolution of $R = 80000$ from $\lambda = 460 - 860$ nm, and uses 3x1 binning along the spatial direction. The data were bias-subtracted, flat-fielded, and extracted using standard IRAF tasks. The reduced spectra were wavelength-calibrated using a ThAr lamp exposure from immediately before the observation of the star. In order to reach high a signal-to-noise ratio and avoid detector saturation, we co-added 7 spectra after reduction and telluric correction (see Section \ref{sec:tellcorr}).

In addition, several M-type stars were observed with the Magellan Inamori Kyocera Echelle (MIKE) optical echelle spectrograph on the Clay telescope at Magellan Observatory as spectral type standards for a study of young stars \citep{Kraus2014}. We used the $0.7''$ slit, which yields a spectral resolution of $R = 35000$ from $\lambda = 335 - 950$ nm. The pixel scale oversamples the resolution with the $0.7''$ slit, so we used 2x binning in the spatial and spectral directions. The spectra were reduced using the CarPy pipeline \citep{Kelson2003}\footnote{\url{http://code.obs.carnegiescience.edu/mike}}. The reduced spectra were wavelength calibrated using a combination of a ThAr lamp exposure and the 760 nm telluric A band \citep[see][for details]{Kraus2014}.

TelFit can handle rapidly rotating early-type stars without preprocessing the spectra by fitting a Savitzky-Golay smoothing filter \citep{savitzky1964} as a pseudo-continuum for the telluric lines. For more feature-rich late-type stars, we remove an approximation to the stellar spectrum using the procedure below before fitting the telluric lines.

\begin{enumerate}
  \item Make an initial guess telluric spectrum with TelFit. The guess spectrum intentionally has slightly higher molecular mixing ratios than the correct values so that it over-corrects the telluric lines.
  \item Divide the observed spectrum by the guess telluric spectrum. This leaves a spectrum with stellar lines in absorption and telluric residuals appearing like emission lines.
  \item Find the best-fit stellar model to the absorption lines using a grid of PHOENIX model spectra \citep{Hauschildt1999} with effective temperature and log(g) near the expected values for the star.
  \item Divide the original data (before the telluric over-correction in step 1) by the best-fit normalized stellar spectrum. This leaves a spectrum that is mostly telluric lines, but which may have strong stellar line residuals.
  \item Enumerate the wavelength ranges that are strongly affected by stellar line residuals, and tell TelFit to ignore them. We ignored all wavelengths from 817.5 - 820 nm in our M-star spectra, which have strong residuals from the sodium lines.
\end{enumerate}

The fit to the primary star does not need to be perfect. Indeed, if we could perfectly model stellar spectra there would be little reason to observe them! The purpose of the preprocessing described above is to ensure that telluric lines dominate the spectrum before fitting. Any strong residuals from that process can and should be ignored in the telluric fit. While a physically better solution would be to fit the stellar and telluric spectra simultaneously, the method described above will often work well enough  for most purposes and is much faster than a simultaneous fit.

\begin{figure*}
\begin{center}
\subfloat{\includegraphics[width=0.9\columnwidth]{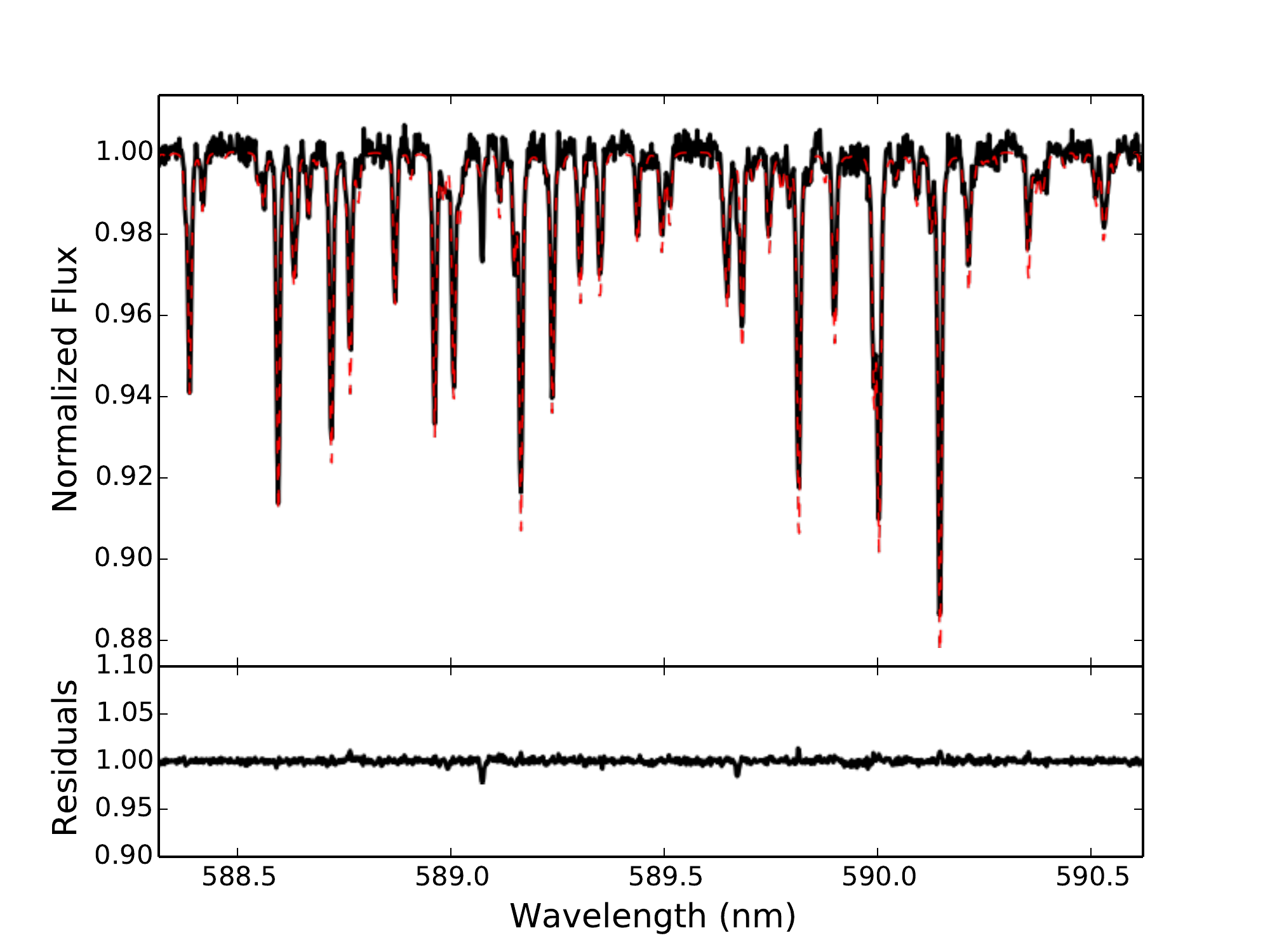}} 
\subfloat{\includegraphics[width=0.9\columnwidth]{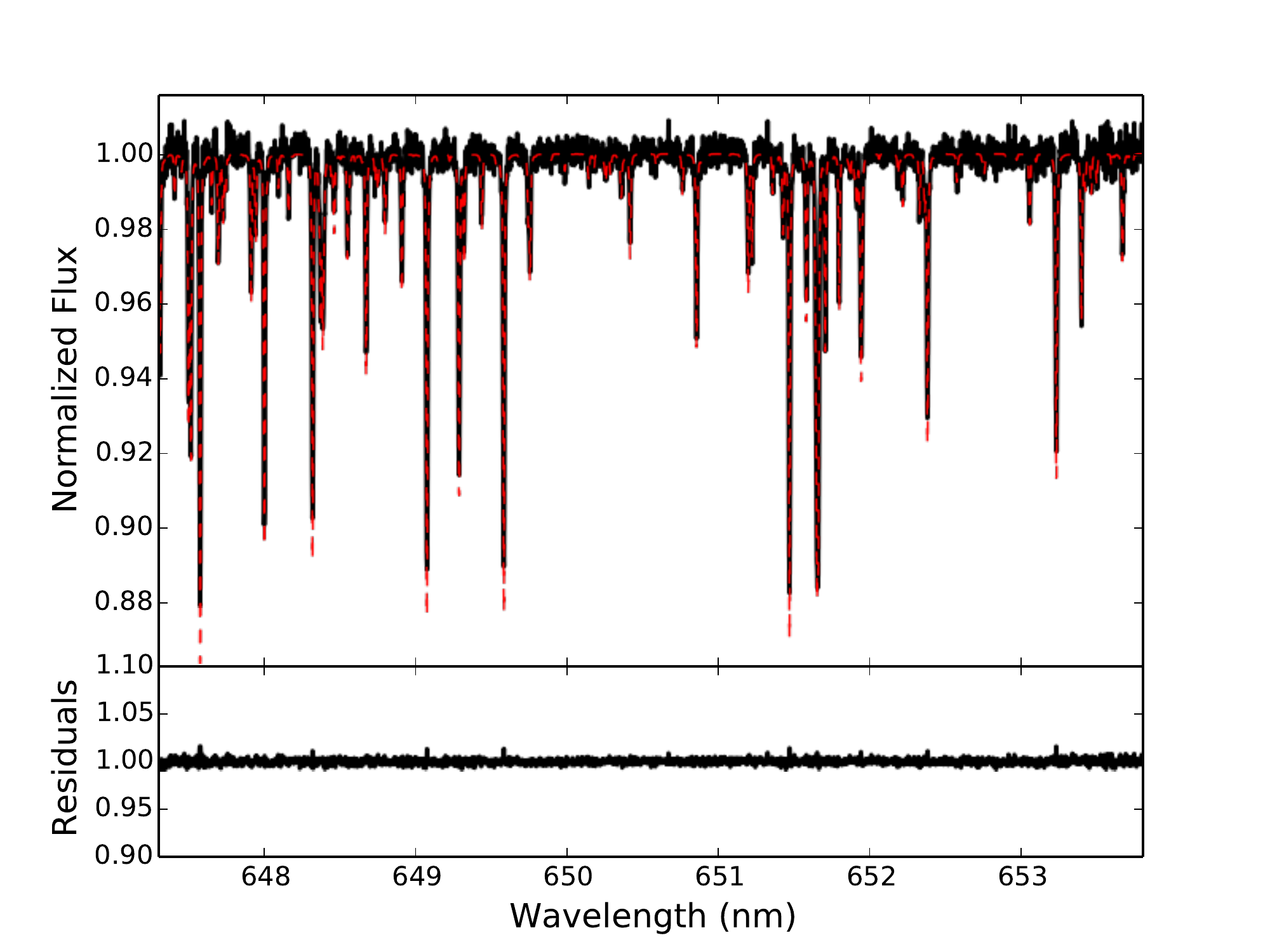}}
\\
\noindent 
\subfloat{\includegraphics[width=0.9\columnwidth]{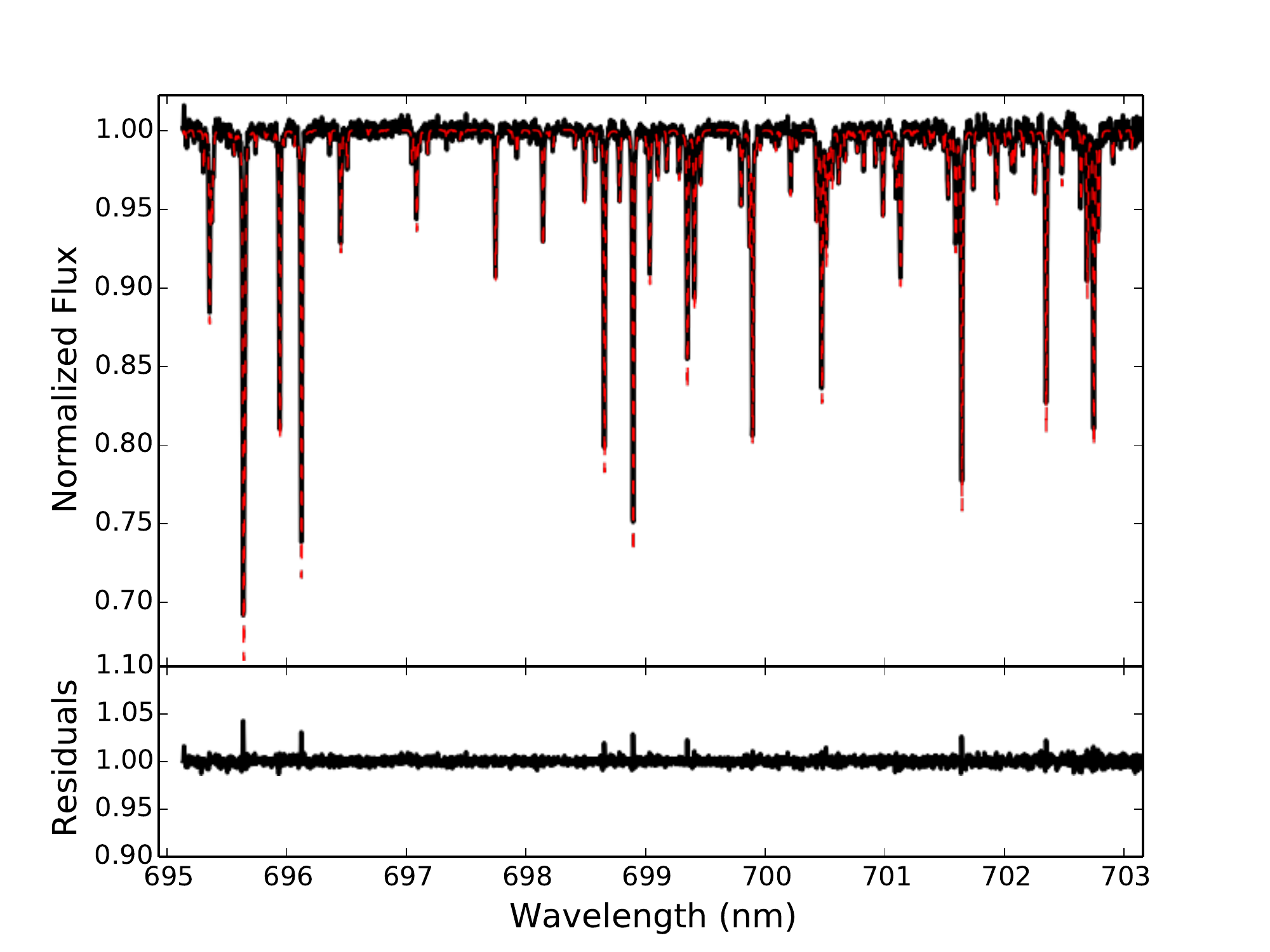}} 
\subfloat{\includegraphics[width=0.9\columnwidth]{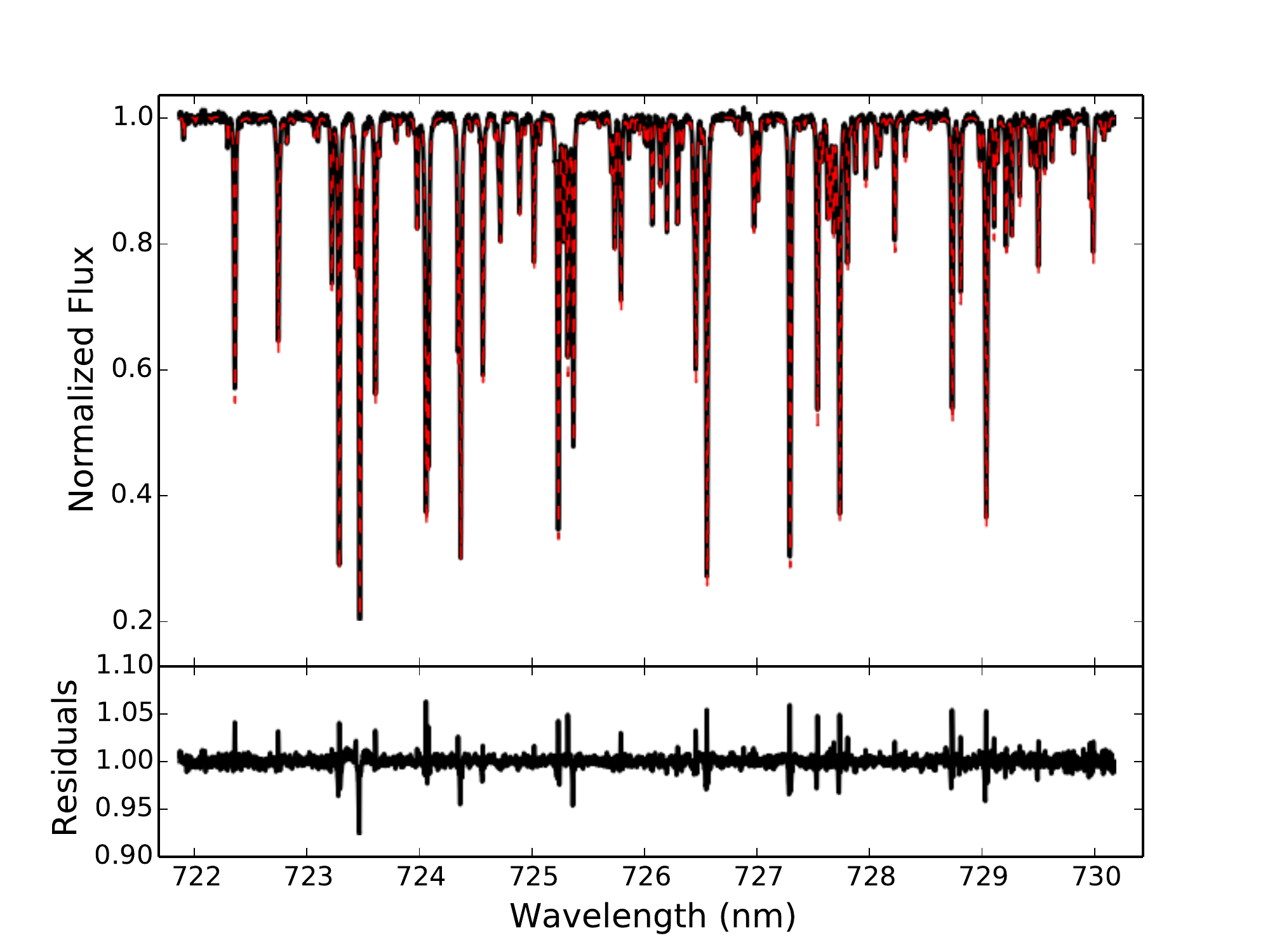}}
\caption{Correction of the water bands in optical spectra. All spectra are of the A0V star HIP 20264, and are smoothed with a Savitzky-Golay filter after telluric correction to remove any broad features in the stellar spectrum. The top panel of each figure shows the observed spectrum (black solid line) and the best-fit telluric model (red dashed line), and the bottom panel shows the residuals after division by the telluric model. The telluric water lines are corrected to very near the noise level of the spectrum in the top row, revealing weak interstellar Na D lines (top left). The telluric correction leaves residuals on the order of $5\%$ of the continuum for strong telluric lines (bottom right), possibly due to an incorrect atmosphere profile for water vapor.}

\label{fig:bstarcorr_water}
\end{center}
\end{figure*}

\section{Telluric Fitting Method}
\label{sec:tellcorr}

The TelFit code performs a least-squares fit using a constrained Levenberg-Marquardt algorithm \citep{Marquardt1963}. It is extremely flexible; the user can decide which atmospheric parameters to fit, which spectral regions to use in the fit, how the detector point spread function (PSF) is modeled, the order of the continuum fit used in each iteration, and whether the data wavelengths are calibrated to fit the model or vice versa. The algorithm will optimize the fitting variables, as well as the detector resolution and the wavelength solution. LBLRTM requires an atmosphere profile giving the temperature, pressure, and the abundance of several molecules as a function of height. We provide a default atmosphere profile\footnote{The default atmosphere profile was developed for the Michelson Interferometer for Passive Atmospheric Sounding (MIPAS) spacecraft, are is available from \url{http://www-atm.physics.ox.ac.uk/RFM/atm/}} with the TelFit code that is acceptable for mid-latitude observatories, but provide the ability to easily give an alternate atmosphere profile for more accurate results. 

While TelFit explicitly fits the temperature, pressure, and abundances at the observatory altitude, it scales the quantities at \emph{all} atmosphere layers. For the pressure and temperature, it finds the difference between the requested pressure (temperature) at the requested altitude, and the pressure (temperature) in the atmosphere profile. If $P_r$ is the requested pressure at altitude $z$, and the atmosphere profile pressure is $P_0 $ at that altitude, then the pressure $P_i $ at each atmosphere layer $z_i$ is scaled as
\begin{equation*}
  P_i \rightarrow P_i - (P_r - P_0) e^{\frac{-(z_i - z)^2}{2\cdot (10 \rm km)^2}}
\end{equation*}
The temperature is scaled in the same way as the pressure above, and is done in this way so that the quantities more than about 10 km above the observatory are effectively unchanged. The mixing ratio of all molecules is scaled in a much simpler way: if $A_r$ is the requested mixing ratio at the telescope altitude, and $A_0$ is the mixing ratio in the atmosphere profile, the mixing ratio at each layer $A_i$ is given by
\begin{equation*}
  A_i \rightarrow A_i  \frac{A_r}{A_0}
\end{equation*}
\emph{Because of the way TelFit scales the temperature, pressure, and molecular mixing ratios in every atmosphere layer, the optimized values should not be taken as a measurement of the actual surface mixing ratios.}

In each iteration for the main variables (temperature, pressure, telescope zenith angle, and molecular abundances), TelFit refines the wavelength solution and detector resolution. The wavelength fitter performs a third-order polynomial fit to adjust the wavelengths of the model such that the telluric lines match the data. Alternatively, the data wavelengths can be adjusted to fit the model. The wavelength solution of the data should be very close, as TelFit will not shift any wavelengths by more than 0.1 nm. By default, TelFit fits the spectrograph PSF as a gaussian, and fits the detector resolution by finding the best gaussian width to convolve the model against. Alternatively, TelFit can do a nonparametric fit to the spectrograph PSF using the singular value decomposition described in detail in \cite{Rucinski1999}

\section{Results}
\label{sec:results}

We fit the spectra for the rapidly-rotating A0V star HIP 20264 with a two-step approach. In the first step, we fit the observatory temperature and humidity in the echelle orders dominated by the water bands at 590 nm, 650 nm, 700 nm, and 730 nm. For the rest of the fit, the humidity and temperature were fixed at the $\chi^2$-weighted average of the fits for each band. Next, we fit the $O_2$ abundance with the $\gamma$ and B bands near 630 nm and 690 nm, respectively. Because the blue end of the B band is extremely strong and absorbs nearly all the light, we only use the red half in the fit. The $O_2$ abundance was fixed at the $\chi^2$-weighted average of the individual fitted values for the two echelle orders. We then applied the best-fit parameters to every echelle order, allowing the fitter to adjust the model wavelengths and detector resolution separately for each order. For all telluric fits in this paper, we adjusted the temperature, pressure, and water vapor atmosphere profile with sounding data from the Global Data Assimilation System (GDAS) meteorological archive\footnote{The GDAS archive is available starting in December 2004 at \url{http://ready.arl.noaa.gov/READYamet.php}. Instructions for its use are included in the code documentation.}. As stated in Section \ref{sec:obs}, we fit each frame of HIP 20264 separately, and co-added the telluric-corrected spectra. The fitted value of the relative humidity varied from $18.7\%$ to $22.1\%$, and the airmass of the star increased from 1.15 to 1.52 (a change of $19^{\circ}$ in zenith angle) over the course of the 7 frames, making the individual fits crucial.

\begin{figure}
\begin{center}
\subfloat{\includegraphics[width= 0.9\columnwidth]{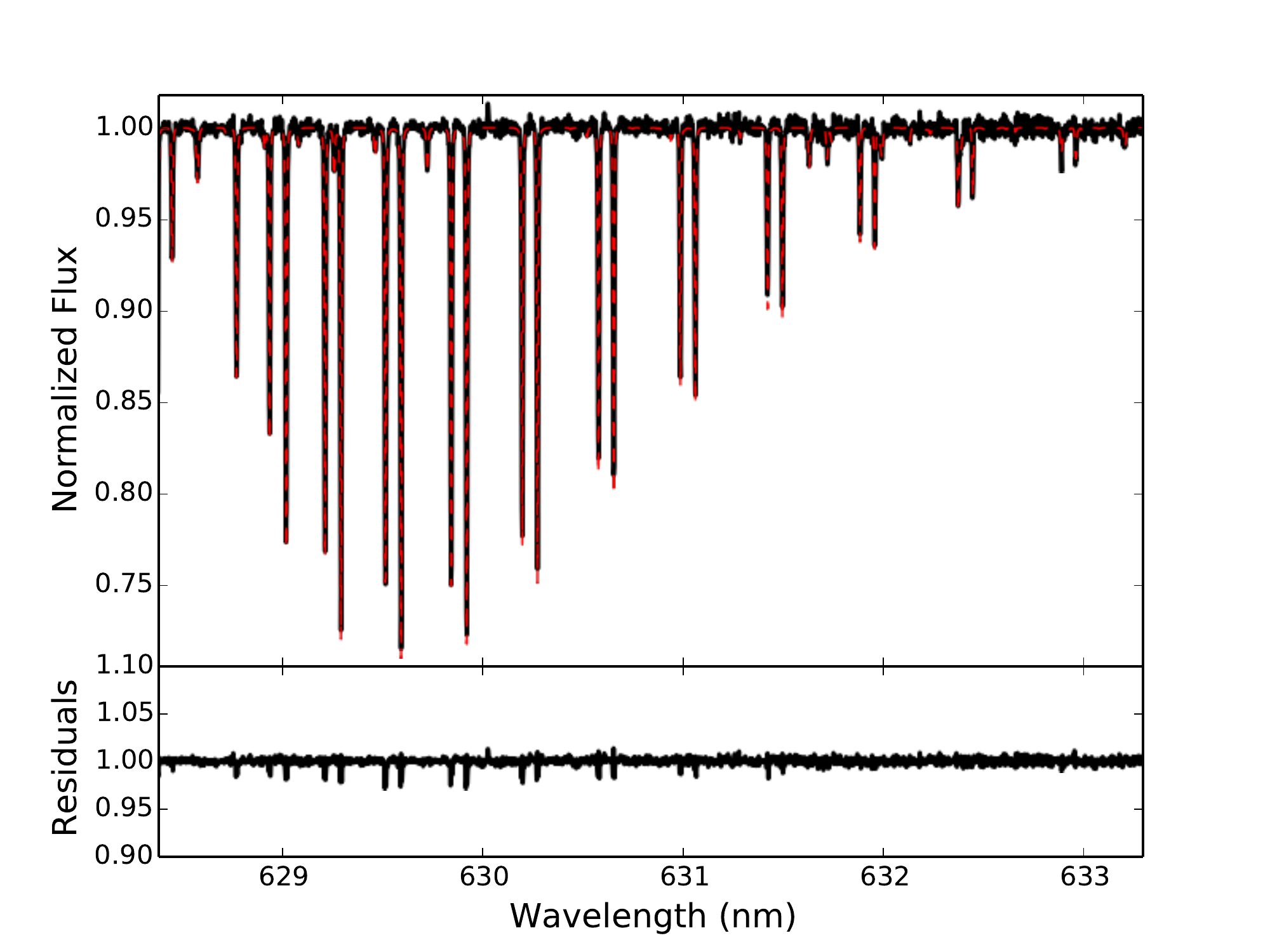}}
\\ 
\subfloat{\includegraphics[width=0.9 \columnwidth]{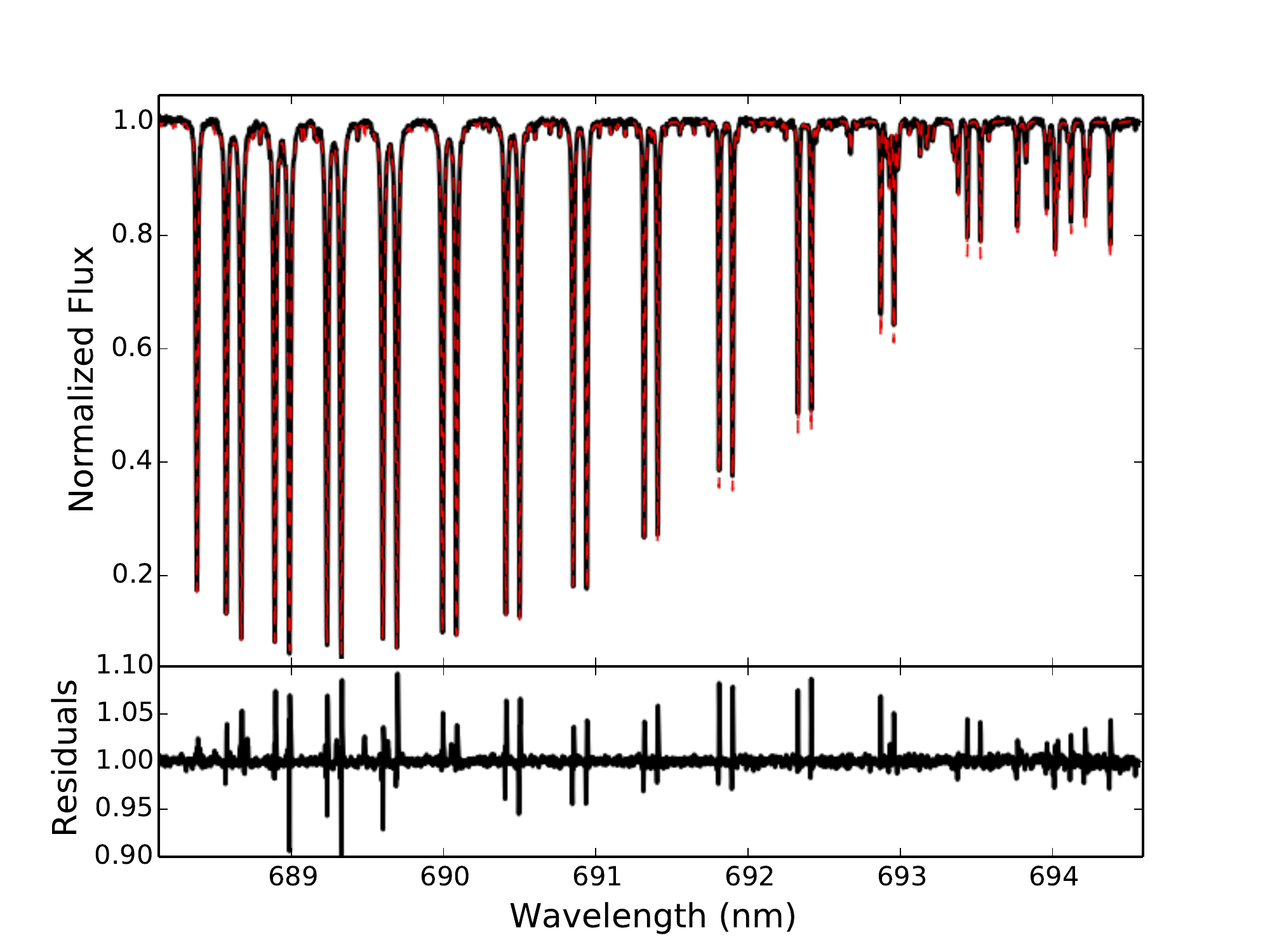}}
\caption{Correction of the $\rm O_2$ bands in optical spectra. All spectra are of the A0V star HIP 20264, and are smoothed with a Savitzky-Golay filter after telluric correction to remove any broad features in the stellar spectrum. The top panel of each figure shows the observed spectrum (black solid line) and the best-fit telluric model (red dashed line), and the bottom panel shows the residuals after division by the telluric model. }

\label{fig:bstarcorr_oxygen}
\end{center}
\end{figure}

We show the telluric correction for the observation of the A0V star HIP 20264 in four water bands and two $\rm O_2$ bands in Figures \ref{fig:bstarcorr_water} and \ref{fig:bstarcorr_oxygen}, respectively. The telluric model shown is the average telluric model of the 7 individual exposures, since we corrected each one separately to better account for the changing water vapor content and telescope zenith angle. We apply a Savitzky-Golay smoothing filter to the corrected data to remove any broad features in the stellar spectrum. The water features are corrected to near the noise level of the spectrum, with the exception of the very strong telluric water band near 730 nm. The poor correction may be due to a slightly incorrect temperature and water-vapor atmospheric profile, which causes the line profile to change appreciably for strong lines. In addition, small errors in the line strength parameters are more noticable for strong telluric lines. The correction in the $\rm O_2$ B and $\gamma$ bands (Figure \ref{fig:bstarcorr_oxygen}) is somewhat worse. The same $\rm O_2$ mixing ratio undercorrects the $\gamma$ band and slightly overcorrects the B band. This systematic error may be from incorrect line strengths in the HITRAN database for the two bands.


We observed the A1V star HIP 25608 30 minutes after HIP 20264 and at similar airmass, and so use it as a telluric standard star to directly compare our method to the empirical telluric correction method. To have comparable S/N ratios as in Figures \ref{fig:bstarcorr_water} and \ref{fig:bstarcorr_oxygen}, we co-added all 7 frames of both the the target star (HIP 20264) and the standard star (HIP 25608). We account for the different column density of absorbers in the target and standard star observations using Beer's law \citep{Beer1852} as follows. For each order, we scaled the normalized flux of the standard star ($f_s$) with

\begin{equation}
f_s \rightarrow f_s^{\gamma}
\label{eqn:correction}
\end{equation}

where $\gamma$ is determined using the normalized flux of the standard star and target star ($f_t$) and is defined as

\begin{equation}
\gamma = \text{median} \left( \frac{\log{f_t}}{\log{f_s}} \right)
\label{eqn:scale}
\end{equation}

\begin{figure*}
\begin{center}
\subfloat{\includegraphics[width=0.9\columnwidth]{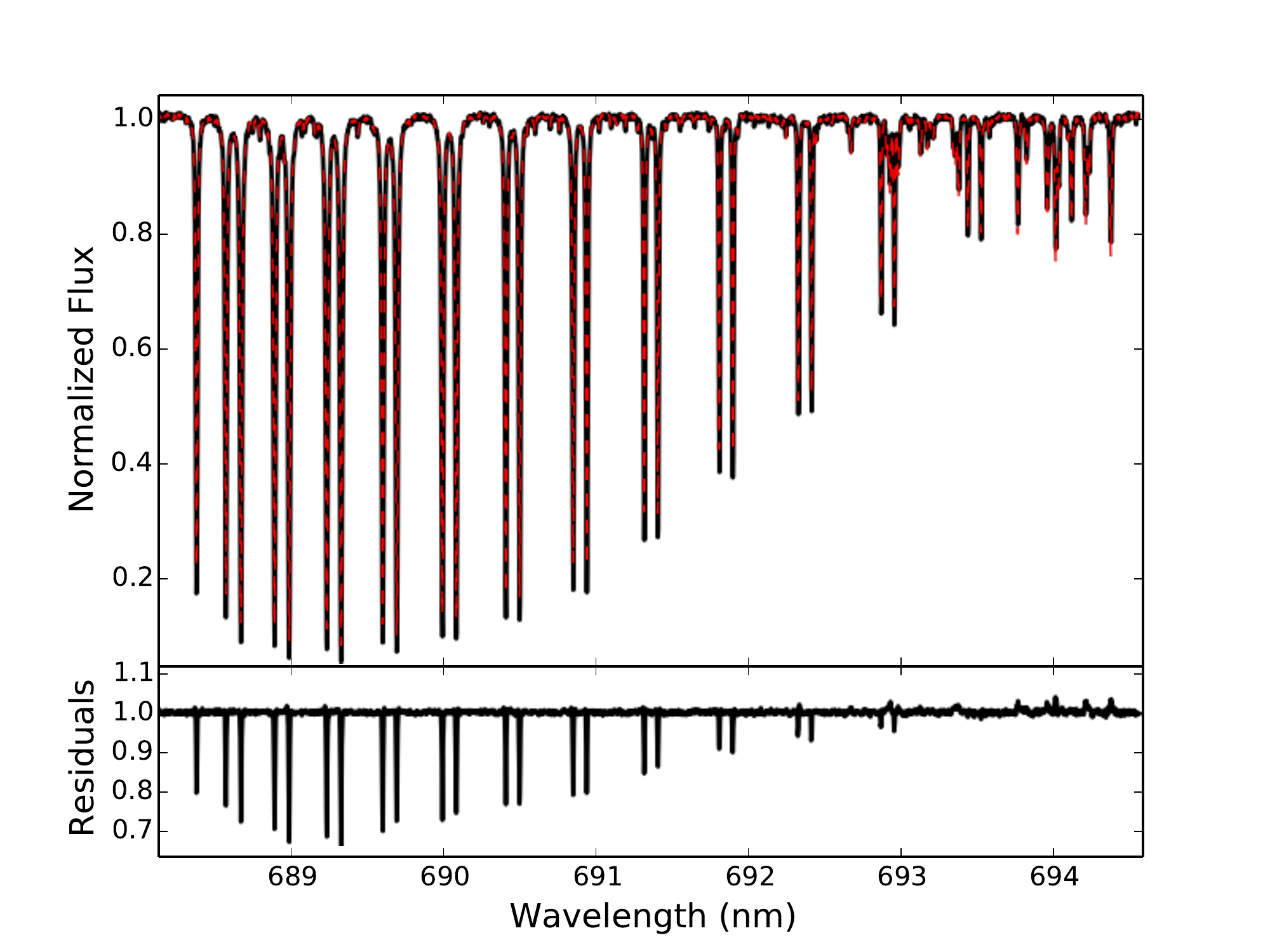}}
\subfloat{\includegraphics[width=0.9\columnwidth]{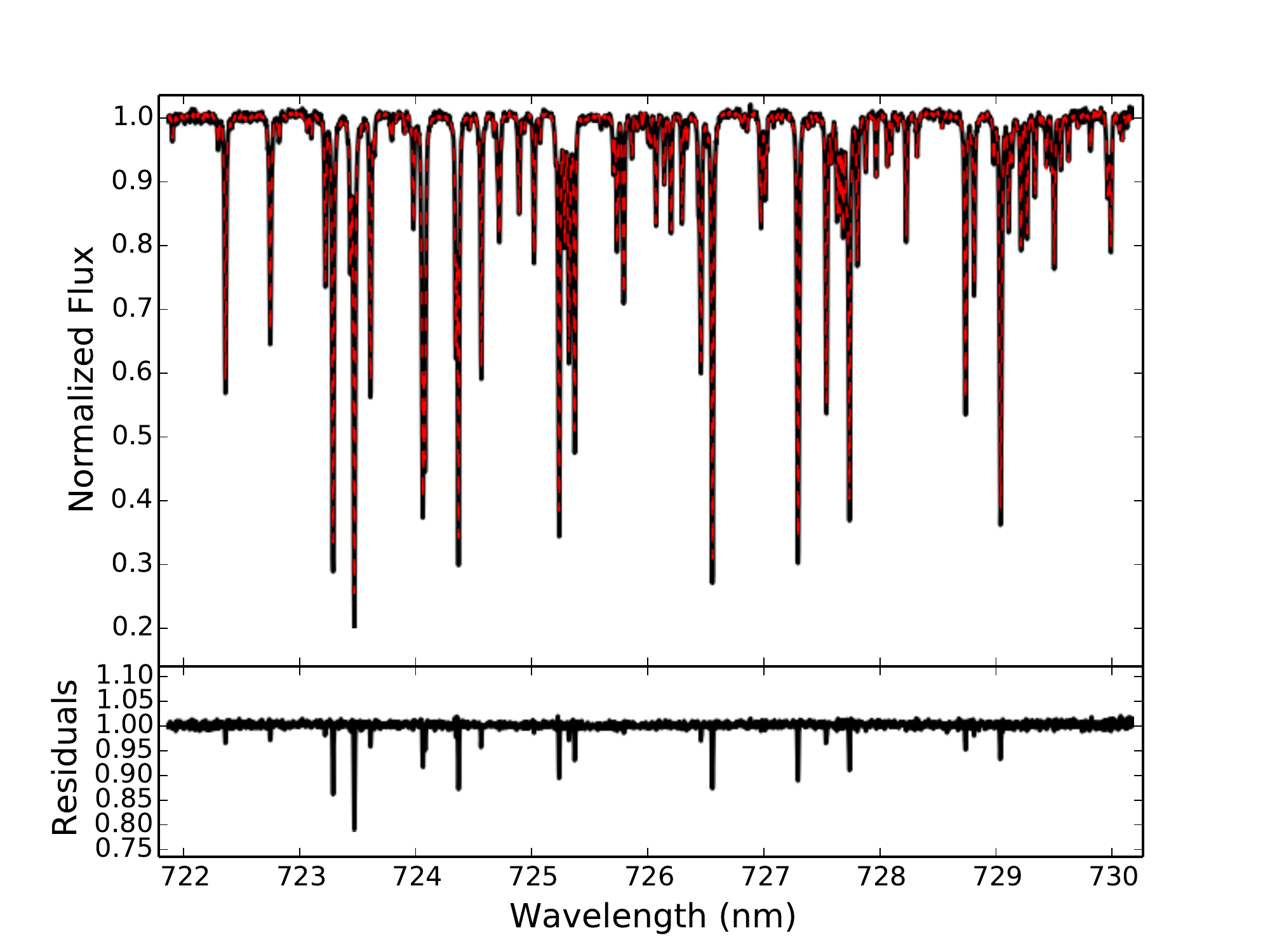}}
\\ 
\subfloat{\includegraphics[width=0.9\columnwidth]{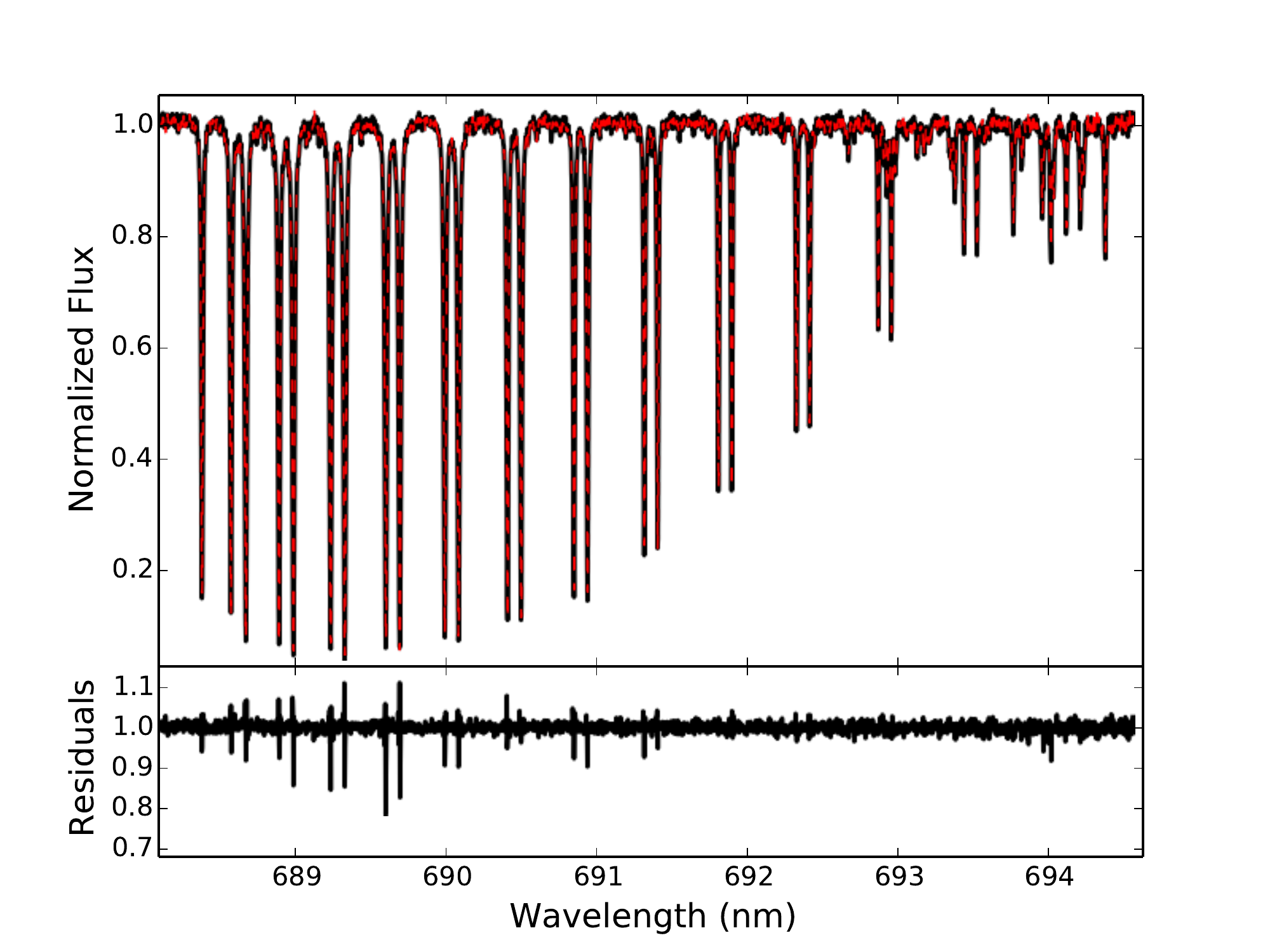}}
\subfloat{\includegraphics[width=0.9\columnwidth]{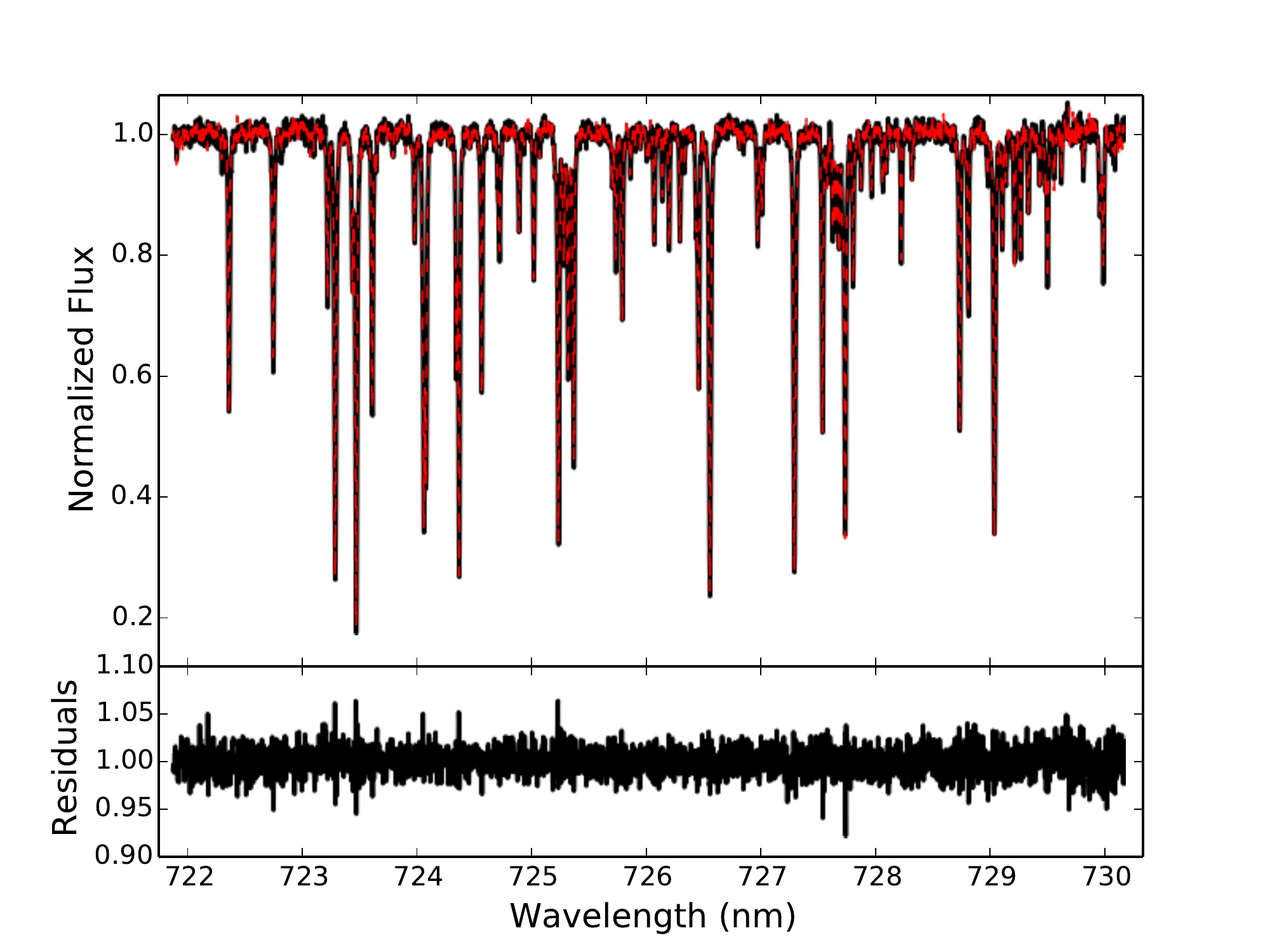}}
\caption{Empirical telluric corrections. \emph{Top Row}: Correction at high S/N ratio, where all 7 frames of both the target star (HIP 20264, A0V) and the telluric standard star (HIP 25608, A1V) were co-added before the telluric correction. In this case, both the humidity and the airmass are changing throughout both exposures and the empirical telluric correction is very poor. \emph{Bottom Row}: Correction between the last frame taken of the target star and the first frame of the standard star, a more common mode of empirical correction. In this case, the empirical telluric correction is comparable to the model fitting method presented in this paper. The left column should be compared to the bottom panel in Figure \ref{fig:bstarcorr_oxygen}, and the right column should be compared to the bottom right subfigure in Figure \ref{fig:bstarcorr_water}. The poor correction in the oxygen band (lower left panel) is the result of the slight airmass difference between the science and standard stars. The correction in equation \ref{eqn:correction} is only valid for weak or moderate lines, and does not work as well for the nearly saturated lines shown here. The empirical correction under-corrects the line core while over-correcting the wings.}

\label{fig:bstarcorr_empirical}
\end{center}
\end{figure*}

In Equation \ref{eqn:scale}, we only use the points with $f_t < 0.95$ in the median computation, so that the noise does not dominate in the orders with few telluric lines. These empirical telluric corrections are shown in the top row of Figure \ref{fig:bstarcorr_empirical}. The corrections are quite poor in this case, mostly because the airmass of the target (standard) star changes from 1.15-1.52 (1.31-1.86) from the first to last frame, and co-adding the spectra amounts to a flux-weighted average over airmass that the correction in Equation \ref{eqn:correction} cannot capture.


We also test an empirical telluric correction using only the last frame of the target star and the first frame of the standard star, and show the result in the bottom row of Figure \ref{fig:bstarcorr_empirical}. In this case, the airmass can be treated as approximately constant and Equation \ref{eqn:correction} does a much better job in accounting for the small airmass difference between the target and standard star spectra, with the exception of the nearly saturated oxygen lines (bottom left panel). The water line correction results in similar line residual amplitudes to the telluric modeling method (see Figure \ref{fig:bstarcorr_water} for comparison).


\begin{figure}[ht]
  \centering
  \subfloat{\includegraphics[width=0.9\columnwidth]{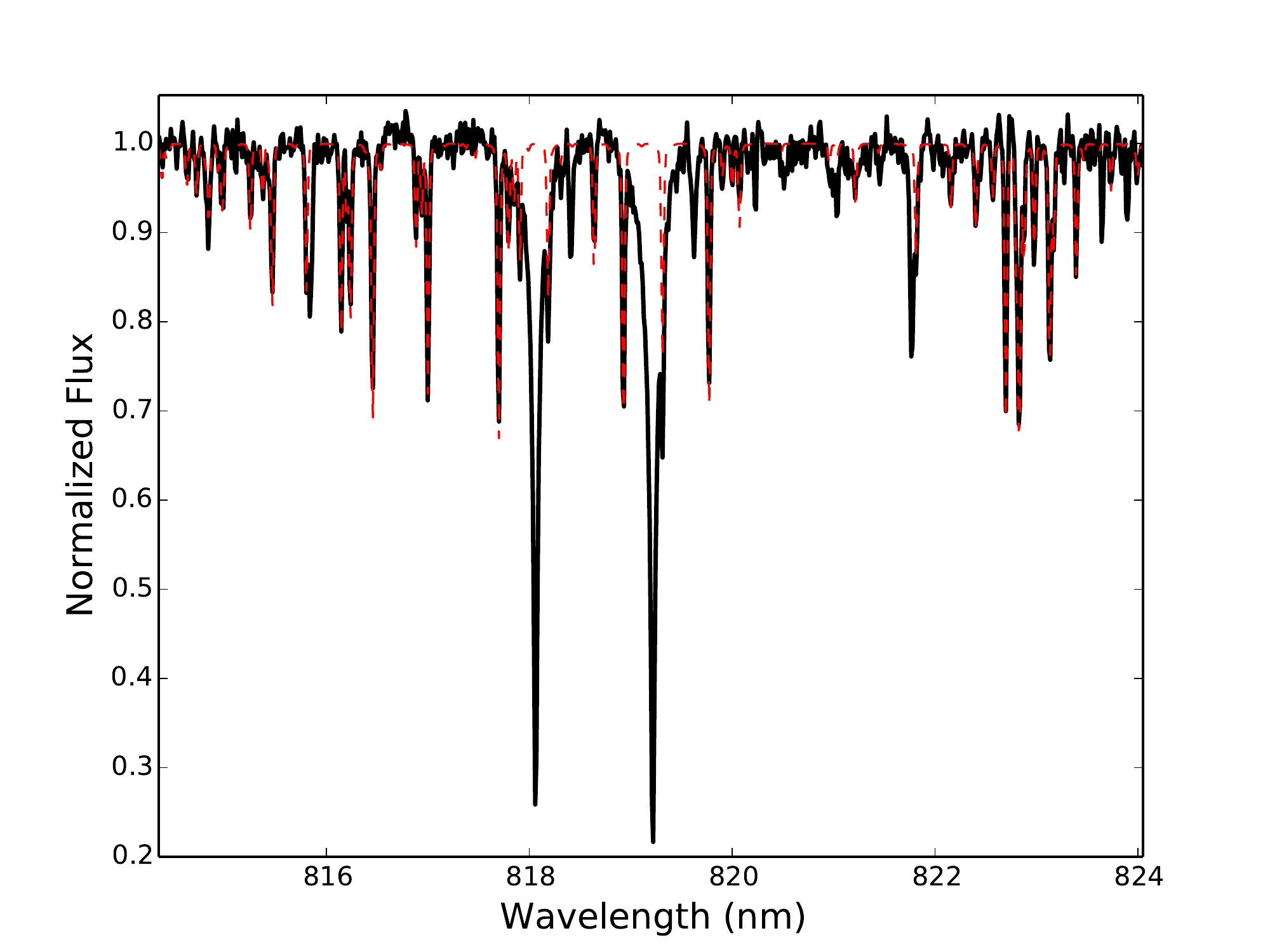}}
  \\
  \subfloat{\includegraphics[width=0.9\columnwidth]{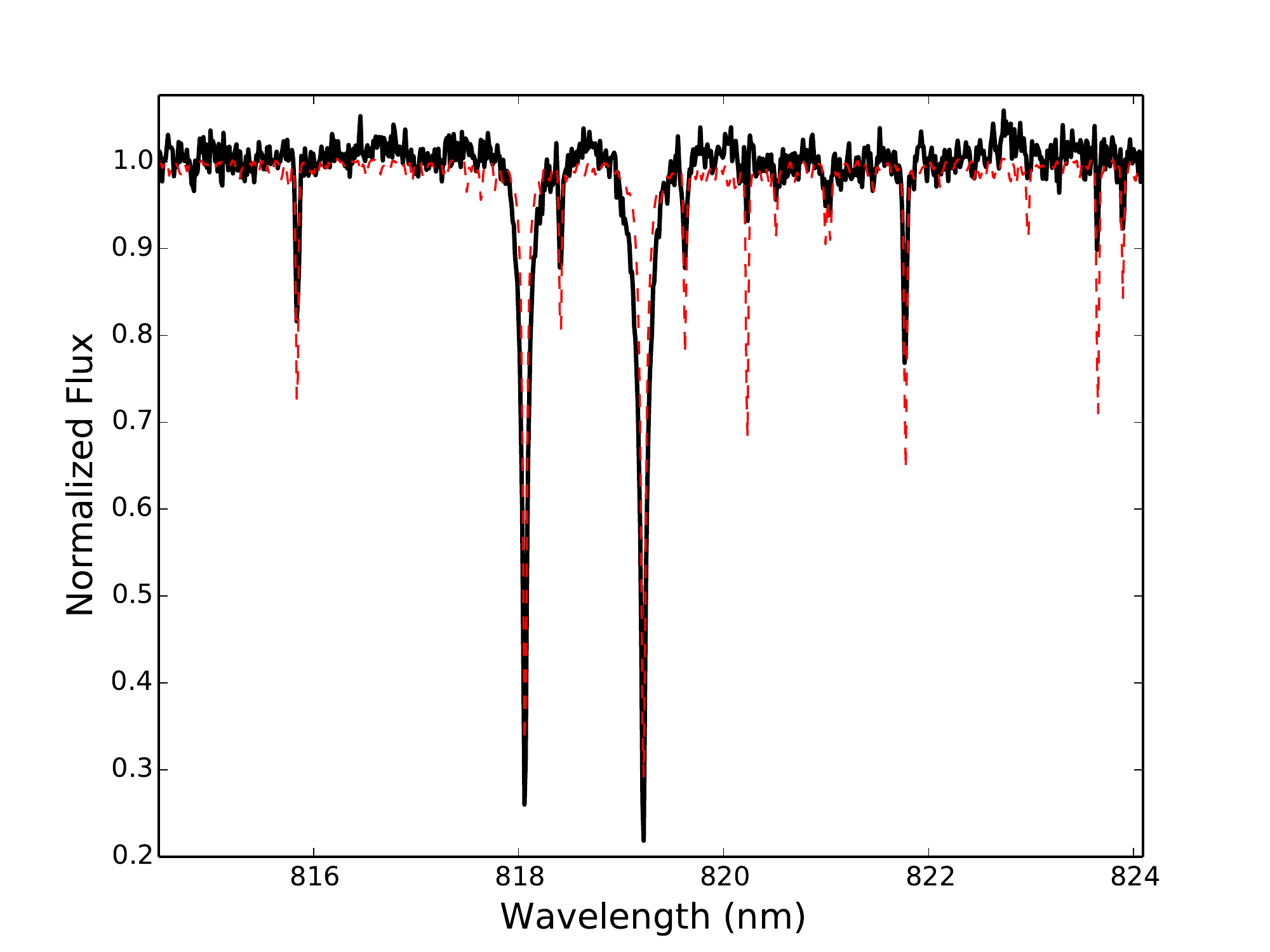}}
  \caption{\emph{Top}: The observed spectrum of GJ 908 (black solid line) with the best-fit telluric spectrum (red dashed line) overplotted. There are telluric lines within the sodium doublet line profiles which affect any line shape measurements if they are not corrected. \emph{Bottom}: The telluric-corrected spectrum of GJ 908 (black solid line) with a PHOENIX model spectrum (red dashed line) overplotted to guide the eye. The model spectrum has the following parameters: $\rm Teff = 3700 K$, $\log g = 4.0$, and $\rm [Fe/H] = -0.5$, and has been convolved with a gaussian to match the detector resolution. All of the remaining absorption lines in the spectrum come from the star.}
  
  \label{fig:mtellcorr}
\end{figure}

We now turn to a more typical-use case for the TelFit code: correcting the telluric lines in late-type stars near a feature of interest. For this, we use a series of M-type stars (see Table \ref{tab:sample}) with observations near the 819 nm sodium doublet, which is used as a gravity-sensitive age indicator for late-type stars \citep{Slesnick2006}. As the stellar age and therefore surface gravity increases, the line strength increases. Amongst main sequence M-stars, later spectral types have higher surface gravity and so we expect the equivalent width of the sodium lines to increase as we go towards later spectral types.

\begin{figure}[ht]
  \centering
  \includegraphics[width=0.9\columnwidth]{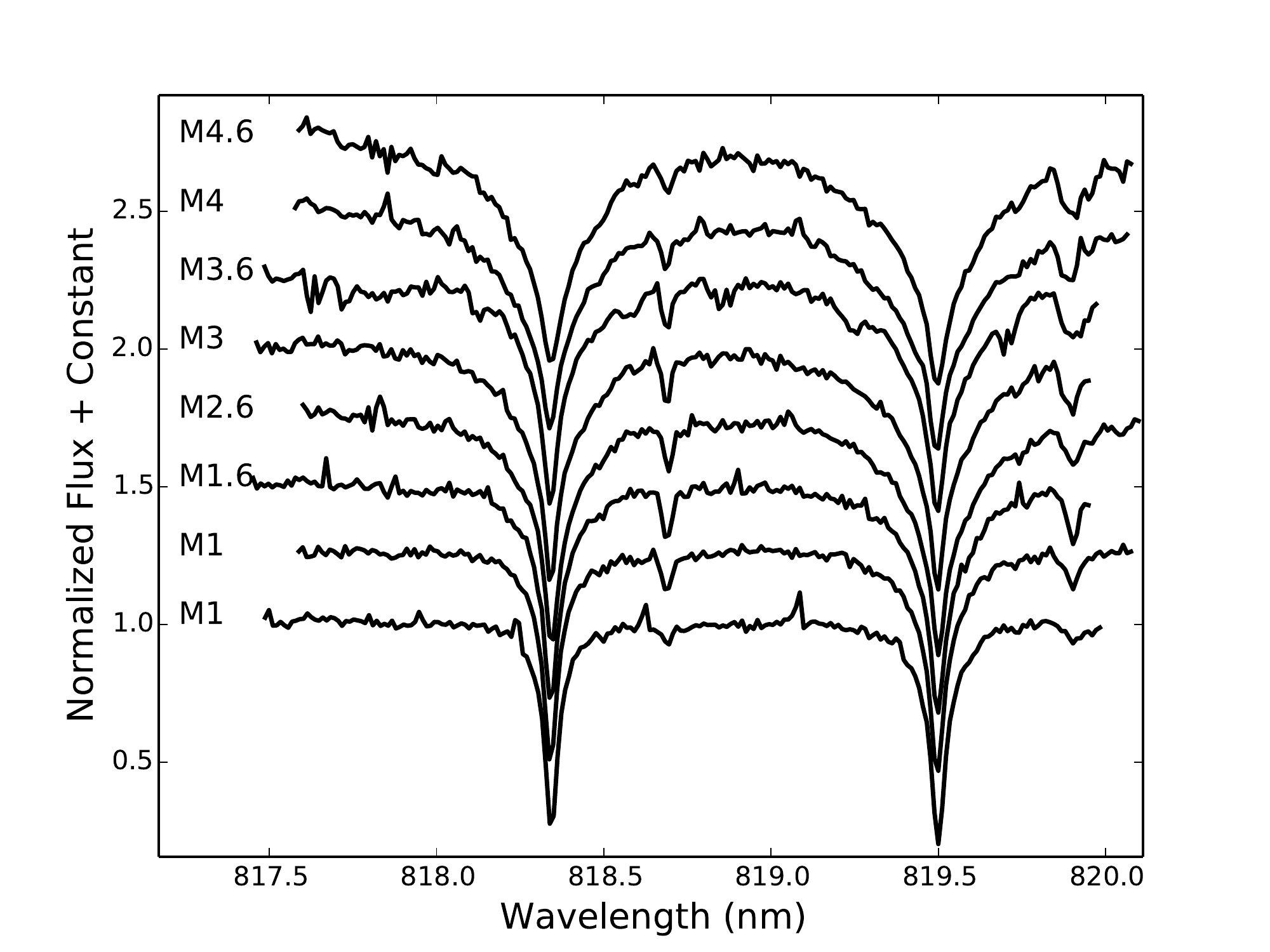}
  \caption{The evolution of the sodium doublet line profile with spectral type for a series of main-sequence M-type stars. The later spectral types show significantly broader line wings.}
  \label{fig:lineprofile}
\end{figure}

Figure \ref{fig:mtellcorr} shows the telluric correction for the M1V star GJ 908. For this and all M-star spectra in this work, we fit the water vapor, temperature, and $O_2$ mixing ratio at the same time for the spectral order covering 819 nm. The bottom panel of Figure \ref{fig:mtellcorr} shows that each of the spectral lines after telluric correction come from the star itself, and that the telluric contamination is reduced to near the noise level of the spectrum. Importantly, the telluric lines that fall within the sodium doublet line profile no longer affect the profile shape. We can now directly compare the line profile of the sodium doublet lines as a function of spectral type, without the influence of telluric lines. Figure \ref{fig:lineprofile} shows the evolution of the doublet lines. The central depth stays about the same throughout the sequence, but the later spectral type stars show significantly broader line wings, leading to the known sequence in equivalent width \citep{Slesnick2006}. The robust recovery of the sodium line strengths and profiles demonstrate that TelFit can accurately remove telluric lines, even in feature-rich spectra.


\section{Conclusions}
\label{sec:conclusion}
We presented the TelFit code, an object-oriented Python code capable of accurately fitting the telluric spectrum in ground-based spectra. We use a high signal-to-noise ratio echelle spectrum of the A0V star HIP 20264 to demonstrate the fit quality in Figures \ref{fig:bstarcorr_water} and \ref{fig:bstarcorr_oxygen}. Contrary to expectation, the water lines are typically fit to much higher precision than the $\rm O_2$ telluric lines. This is likely coming from a systematic error in the HITRAN line strength database, since the same $O_2$ mixing ratio overfits the B band and underfits the $\gamma$ band. We compare our code to an empirical telluric correction of HIP 20264, and find that the model is at least as accurate. In fact, TelFit is significantly more accurate than the empirical method when several frames of both the target and standard star are co-added.

We also demonstrated the use of TelFit for in-depth analysis of spectral features in late-type stars with a series of M-star observations near the 819 nm sodium doublet. The telluric lines were removed to near the noise level of the observations, allowing for analysis of the sodium line profiles and recovery of the known sequence of increasing equivalent width with later spectral types. Regions contaminated by telluric lines are often ignored in optical spectral analysis; accurate correction of telluric features could help open these regions up for further analysis.

This code was mostly developed and tested for correction of optical spectra. However, an early version of this code was used in \cite{Gullikson2013} to correct for telluric methane absorption in B-star spectra, with similar line residuals to those that \cite{Seifahrt2011} found in the same spectral region. We encourage the use of TelFit for correcting telluric absorption in near-infrared as well as optical data.

We would like to thank Andreas Seifahrt for his help in the early stages of code development. This project was funded by a UT Austin Hutchinson fellowship to Kevin Gullikson and start-up funding to Sarah Dodson-Robinson from the University of Texas. This work makes use of Astropy, a community-developed core Python package for Astronomy \citep{Astropy}, as well as SciPy and NumPy \citep{Oliphant2007}. We would like to thank the developers of all of those packages.

\end{document}